\documentclass[conference]{IEEEtran}
\usepackage{graphicx}
\usepackage{amsmath, amssymb}
\usepackage{bbm}
\usepackage{cite}
\usepackage{mathtools}
\usepackage{bbold}
\usepackage{dsfont}
\usepackage{amsfonts}
\usepackage{tikz, authblk}
\usetikzlibrary{arrows,automata,positioning}

\begin{document}

\title{Timeliness in NextG Spectrum Sharing under Jamming Attacks with Deep Learning}

\author{\normalsize Maice Costa and Yalin E. Sagduyu \\
Nexcepta, Gaithersburg, MD, USA \\ 
\{mcosta, ysagduyu\}@nexcepta.com
}

\maketitle

\begin{abstract}
We consider the communication of time-sensitive information in NextG spectrum sharing where a deep learning-based classifier is used to identify transmission attempts. While the transmitter seeks for opportunities to use the spectrum without causing interference to an incumbent user, an adversary uses another deep learning classifier to detect and jam the signals, subject to an average power budget. We consider timeliness objectives of NextG communications and study the Age of Information (AoI) under different scenarios of spectrum sharing and jamming, analyzing the effect of transmit control, transmit probability, and channel utilization subject to wireless channel and jamming effects. The resulting signal-to-noise-plus-interference (SINR) determines the success of spectrum sharing, but also affects the accuracy of the adversary's detection, making it more likely for the jammer to successfully identify and jam the communication. Our results illustrate the benefits of spectrum sharing for anti-jamming by exemplifying how a limited-power adversary is motivated to decrease its jamming power as the channel occupancy rises in NextG spectrum sharing with timeliness objectives.
\end{abstract}


\section{Introduction}
The increasing demand for spectrum resources has led to a critical need for innovative strategies to address the challenges posed by spectrum scarcity \cite{matinmikko2020spectrum, gur2020expansive, yang2021spectrum}. As traditional frequency bands become congested and the spectrum becomes a valuable and limited commodity, it has been imperative to explore novel approaches to optimize spectrum utilization. One promising avenue is spectrum sharing, where multiple transmitters coexist within the same frequency band, aiming to enhance overall spectral efficiency. To that end, an emerging example is spectrum sharing in the 3.5GHz Citizens Broadband Radio Service (CBRS) band, where the spectrum band needs to be shared with the radar that serves as the incumbent user.  

The urgency of addressing spectrum scarcity is further underscored by the growing importance of timely communication in various applications, ranging from mission-critical first responder and disaster recovery operations to emerging technologies like the Augmented Reality (AR), Virtual Reality (VR), Extended Reality (XR), Vehicle-to-Everything (V2X), Internet of Things (IoT), and smart cities.  Information freshness in timely communications is essential in NextG applications where real-time data dissemination is of paramount importance \cite{popovski2022perspective}.
For example, V2X communication, essential for the advancement of smart transportation systems, requires timely exchange of information between vehicles and infrastructure to enhance road safety and traffic efficiency. Similarly, the diverse ecosystem of IoT applications, spanning from smart cities to industrial automation, relies on timely data updates to make informed decisions and take timely actions. 
Age of Information (AoI) denoting the time elapsed since the generation of the last received update provides the mathematical framework to assess the timeliness in information transfer \cite{Yates2012}. A lower AoI indicates that the delivered data is not only current but also aligns seamlessly with the real-time requirements of the emerging NextG applications. This alignment, in turn, amplifies the efficiency and responsiveness of real-time experiences, optimizing decision-making processes and enabling prompt actuation based on the most recent data. 

The pursuit of timeliness in NextG communication is not without its challenges. Given the open and shared nature of wireless communications, NextG is susceptible to new threats and exploits \cite{mao2023security}.
The use of adversarial machine techniques to prevent eavesdroppers from detecting ongoing transmissions has been discussed in \cite{Kim5G, CovertRIS, kim2022adversarial}. 
The outcomes of eavesdropping can be used consequently for jamming the ongoing transmissions. 
To support time-sensitive applications, timely communications has been considered under the objectives of covertness \cite{CovertAoI23, Yang21, Wang21, costa2023timely} and anti-jamming \cite{xiao2018dynamic, garnaev2019maintaining, banerjee2022age}.

This paper studies the intricate interplay between spectrum sharing, timeliness of communications, and the vulnerability to eavesdropping and jamming attacks. 
Our results aim to unravel the dynamics of this complex environment, where the classification performance of the deep learning-based system influences the selection of jamming power by the adversary. 
Recognizing the constraints of an adversary with limited resources, we investigate the average AoI as a key metric to quantify the impact of our proposed approach.
Importantly, we demonstrate the advantages of spectrum sharing, showcasing how an adversary with a restricted power budget is incentivized to reduce its jamming power as the channel occupancy increases.
The main contributions of this paper include:
\begin{itemize}
    \item The analytical characterization of AoI in the proposed scenario with spectrum sharing in the presence of an adversary, considering the detection errors that may occur in spectrum sensing and eavesdropping. 
    \item A discussion about the use of deep-learning based classifiers both by the transmitter and by the adversary to guide the decisions to transmit according to the goals of each party. We include the effects of the wireless channel, as well as design parameters as the packet size. We show that smaller packet sizes decrease detection accuracy, and may result in smaller average AoI in a scenario with scarce spectrum resources and the presence of an adversary. 
    \item A thorough investigation of the effects of channel occupancy on the selected jamming power and resulting average AoI. We show the adjustment of jamming power according to spectrum utilization, and highlight the reduction of the AoI when more spectrum resources are available or less jamming power is used to cause interference. 
\end{itemize}

The remainder of the paper is organized as follows. Sec.~\ref{sec:system_model} describes the system model. Sec.~\ref{sec:analysis} evaluates the performance in terms of average AoI and jamming power in spectrum sharing. 
Sec.~\ref{sec:conclusion} concludes the paper.

\section{System Model} \label{sec:system_model}
\subsection{Communication Model}
\begin{figure}
    \centering
    \includegraphics[width = 0.8\columnwidth]{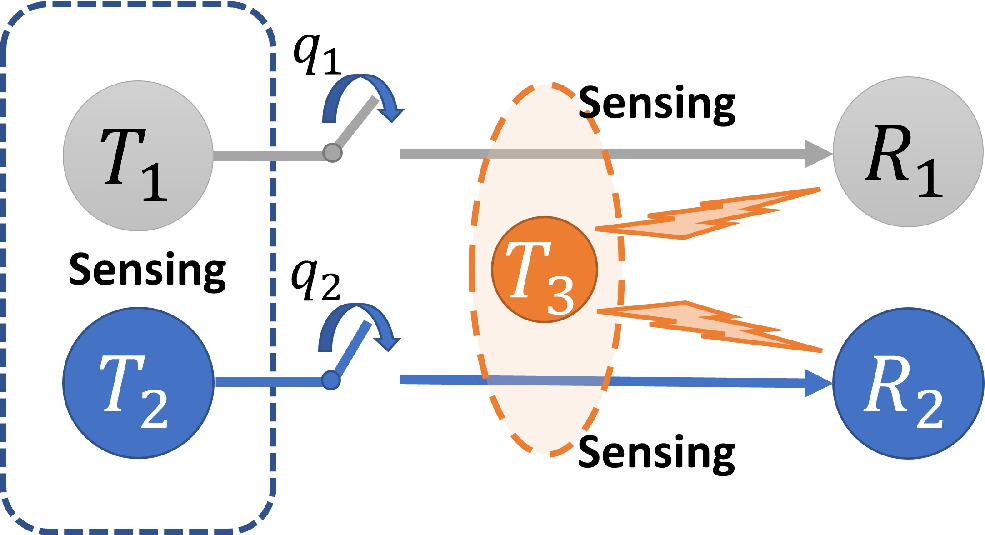}
     \setlength{\belowcaptionskip}{-18pt}
    \caption{Network system model.}
    \label{fig:net}
\end{figure}

We consider a communication network setting with two transmitter-receiver pairs sharing the spectrum resources. We denote with $T_1$ and $T_2$ the transmitter nodes, and assume that $T_1$ (incumbent or primary user) has absolute priority to use the spectrum, while $T_2$ (secondary user) should listen to the channel and transmit only if $T_2$ is silent. Receiver nodes are denoted with $R_1$ and $R_2$ for incumbent and secondary users, respectively. Communication is assumed to take place in a hostile environment where an active adversary ($T_3$) can potentially eavesdrop and jam the signal of both transmitters. The network model is illustrated in \ref{fig:net}.

We assume that at each time slot, $T_1$ will occupy the spectrum with probability $q_1$ independent of other time slots. The secondary user $T_2$ will use a deep learning classifier to decide about the presence of the incumbent at each time slot. This detection task is subject to Type 1 (false positive) and Type 2 (false negative) errors. Misdetection probability is denoted with $p_m$ and false alarm probability is denoted with $p_f$. At a given time slot, we denote with $q$ the probability that the secondary has packets to transmit, but it will act on its intention only if the channel is believed to be available. We denote with $q_2$ the probability of a transmission from $T_2$, where 
\begin{equation}
    q_2 = [(1-q_1)(1-p_f)+q_1p_m]q.
\end{equation}

The eavesdropper $T_3$ also uses a deep learning classifier to decide about channel occupancy. We assume that it will emit an interfering signal if and only if it detects that the channel is occupied (by either transmitter). The set of active nodes impacts the classification accuracy, and we denote the errors with $p_m^{i}$ and $p_f^{i}$ when the transmitter $T_i$ is active, and with $p_m^{12}$ and $p_f^{12}$ when both the transmitters are active. We assume that the secondary transmitter is capable of making a decision before the eavesdropper, so $T_2$ only considers the activity of $T_1$ when deciding to transmit, while $T_3$ considers the activity of both $T_1$ and $T_2$. It is reasonable to expect that $p_m^{12}\leq p_m^{i}$ and $p_f^{12}\leq p_f^{i}$, for $i\in\{1,2\}$, as the SNR increases when combining the transmit powers of both $T_1$ and $T_2$. 

Transmit powers are denoted with $P_i$, $i\in\{1,2,3\}$. We assume that transmissions take place using fixed and independent resource blocks with binary phase shift keying (BPSK) modulation and transmissions are subject to Rayleigh fading plus Gaussian noise of average power $\sigma^2$. We assume that channels between two nodes are independent. The corresponding received power at another node $k$ is $R(i,k) = P_i g(i,k)h(i,k)$, where $g(i,k)$ represents a path gain that may include shadowing and attenuation as a function of distance between nodes, or antenna gain. For the purposes of this work, we assume $g(i,k)$ to remain constant. $h(i,k)$ represents the small scale fading. In the case of Rayleigh fading, the signal envelope follows a Rayleigh distribution, while the received power follows an exponential distribution of parameter $h(i,k)$. 

We assume that a transmission is successful if the signal-to-noise-plus-interference (SINR) is above a given threshold $\gamma_{min}$. Let $\mathcal{A}$ denote the powerset of $\{1,2,3\}$. For a given set $A\in \mathcal{A}$ of active nodes transmitting simultaneously in the same channel (and causing interference to each other), the probability that the intended receiver can successfully decode the signal is given by 
\begin{multline}
    S_i(A) = \exp{-\frac{\gamma_{min}\sigma^2}{P_i g(i,i)h(i,i)}} \times \\
\prod_{j\in A \setminus \{i\}}\left[1+\gamma_{min}\frac{P_j g(j,i)h(j,i)}{P_i g(i,i)h(i,i)}\right].
\end{multline}

The probability of a successful transmission for the pair of $T_i$ and $R_i$, $i\in\{1,2\}$, is calculated averaging over all possible sets of active nodes, 
\begin{equation}
    S_i = \sum_{A\in\mathcal{A}}S_i(A) \mathds{P}(A), 
\end{equation}
where we assume $P_i=0$ if $T_i$ is not active to simplify notation and average over all sets. The term $\mathds{P}(A)$ represents the probability of a given set of active nodes $A$, expressed as a function of the transmit probabilities for each type of node. For example, when the incumbent transmitter $T_1$ is silent, while the secondary transmitter $T_2$ is active and the jammer $T_3$ is active, $A=\{2,3\}$. We express all the events with an active secondary user and the respective probabilities as
\begin{align}
&\mathds{P}(A=\{1,2,3\})=q_1 p_m q (1-p_m^{12}),\\
&\mathds{P}(A=\{2,3\})=(1-q_1)(1-p_f)q (1-p_m^2),\\
&\mathds{P}(A=\{1,2\})=q_1 p_m q p_m^{12},\\
&\mathds{P}(A=\{2\})=(1-q_1)(1-p_f)q p_m^2.
\end{align}

\subsection{Adversary Model}
We assume an adversary ($T_3$) that listens to the communication channel and uses a deep learning classifier to decide about the presence of a signal to interfere with. 
When a signal is detected, $T_3$ actively jams the signal with the objective to increase the interference level and disrupt the communication in the channel. We assume that the adversary cannot distinguish between the incumbent or secondary transmissions, so the decision to jam is triggered by any activity in the channel. We assume that $T_3$ never transmits an interfering signal when it believes the channel is idle, but the output of the classifier is imperfect, with Type 1 (false positive) and Type 2 (false negative) errors. 
The accuracy of the classification task depends on the transmitted signal and the channel quality. We assume that transmit power and/or channel conditions may be different between nodes in the network. We denote with $q_3$ the probability that the jammer will be activated. We express all the events with an active jammer and the respective probabilities as
\begin{align}
   &\mathds{P}(A=\{1,2,3\})=q_1 p_m q (1-p_m^{12}),\\
    &\mathds{P}(A=\{2,3\})=(1-q_1)(1-p_f)q (1-p_m^2),\\
    &\mathds{P}(A=\{1,3\})=q_1 ((1-p_m)+p_m(1-q)) (1-p_m^{1}),\\
    &\mathds{P}(A=\{3\})=(1-q_1)(p_f+(1-p_f)(1-q)) p_f^{12}.
\end{align}
The probability of an active jammer is given by the sum of probabilities of the four events, as
\begin{equation}
    q_3 = \sum_{\{A\in\mathcal{A}:A \supseteq \{3\}\}}  \mathds{P}(A).
\end{equation}

The decision to jam the detected signal is subject to an average jamming power constraint $\bar{P}_{max}$ that represents the concern with a limited power budget. The average power $\bar{P}_3$ satisfies $\bar{P}_3 \leq \bar{P}_{max}$, with $\bar{P}_3 =P_3q_3$.

\subsection{Status Updating Model}
We assume that the communication between $T_i$ and $R_i$ concerns time-sensitive information, so the transmitter obtains the packet immediately before transmission. At the receiver $R_i$, the AoI evolves as
\begin{equation}
 \Delta_i(t+1) =
    \begin{cases}
      1 & \text{w.p. } S_i, \\
      \Delta_i(t)+1 & \text{w.p. } 1-S_i.
    \end{cases}       
\end{equation}

This process can be described as a Discrete Time Markov Chain (DTMC) \cite{pappas22}. At state $k$ the chain can either transition to state $k+1$, when no packet is received, or to state $1$, when a packet is received and AoI is updated. The steady state distribution is $\pi^{\Delta_i}_k = (1-S_i)^{k-1} S_i$, for all $k$. The average AoI is calculated as 
\begin{equation}
\begin{split}
    \bar{\Delta}_i &= \sum_{k=1}^{\infty} k \; \pi^{\Delta_i}_k\\
    &= \sum_{k=1}^{\infty} k (1-S_i)^{k-1} S_i\\
    &= \frac{S_i}{1-S_i} \sum_{k=1}^{\infty} k (1-S_i)^{k}\\
    &=  \frac{S_i}{1-S_i} \frac{1-S_i}{S_i^2}.
\end{split}
\end{equation}
As a result, we have $\bar{\Delta}_i=1/S_i$. 

\section{Performance Analysis}\label{sec:analysis}

\subsection{Signal Detection}
Spectrum data characteristics can be effectively captured by deep learning, providing higher accuracy in wireless signal classification compared to simpler machine learning models or other statistical methods such as energy detection \cite{west2017deep, shi2019deep}. We assume that the secondary user uses a convolutional neural network (CNN) for spectrum sensing, while the adversary uses a feedforward neural network (FNN).  
We consider Glorot uniform initializer, Adam optimizer, and categorical cross entropy loss function to implement a binary classifier with labels `Signal' vs. `No Signal' for these deep learning models. 
We show the deep neural network architectures in Table~\ref{tab:NNArch}.
\begin{table}[b]
    \small
    \centering
    \caption{Deep neural network architectures.}
    \label{tab:NNArch}
    \begin{tabular}{l||l}
     \textbf{FNN}   & \textbf{CNN}  \\
     \hline
       Dense (64, ReLU)  & Conv2D ((1,3), ReLU)\\
       Dropout ($0.1$) & Flatten\\
       Dense (16, ReLU) & Dense (32, ReLU)\\
       Dropout ($0.1$) & Dropout ($0.1$)\\
       Dense (4, ReLU) & Dense(8, ReLU)\\
       Dropout ($0.1$) & Dropout ($0.1$)\\
       Dense (2, SoftMax) & Dense (2, SoftMax)
    \end{tabular}
\end{table} 

\begin{figure}
    \centering
    \vspace{-0.3cm}
    \includegraphics[width=0.95\columnwidth]{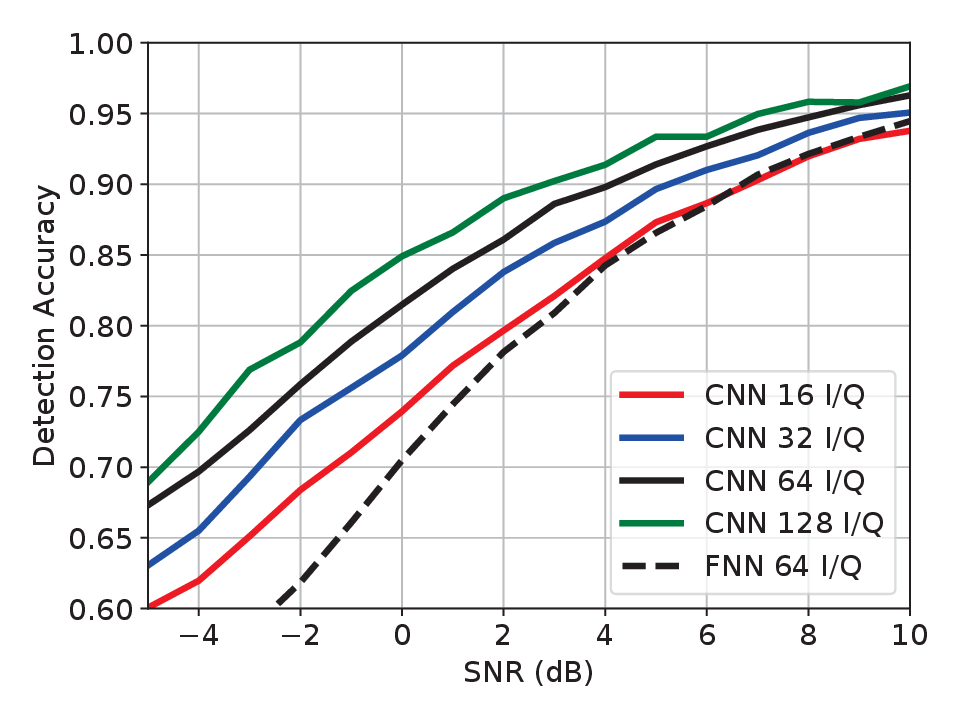}
     \setlength{\belowcaptionskip}{-12pt}
    \caption{Classifier accuracy versus SNR and effect of number of I/Q symbols per packet.}
    \label{fig:acc_symbols}
\end{figure}

Fig.~\ref{fig:acc_symbols} shows the classification accuracy for CNN and FNN models with packets of 64 I/Q samples. Overall, CNN outperforms FNN in terms of classification accuracy that depends on the SNR in the channel between the transmitter and the adversary. We also note that the detection accuracy increases with the packet size as we consider packets with 16, 32, 64 and 128 I/Q samples. The increased accuracy comes at the expense of large number of parameters for the classifier. The number of parameters increases from $3,230$ to $17,566$ for FNN and from $37,306$ to $266,682$ for CNN, when we increase packet size from $16$ to $128$. When fixing the packet size, we adopt the value of 64 I/Q samples, with 9,374 trainable parameters for FNN and 135,610 trainable parameters for CNN. 

\subsection{Communication Timeliness}
\begin{figure}
    \centering
    \includegraphics[width=0.95\columnwidth]{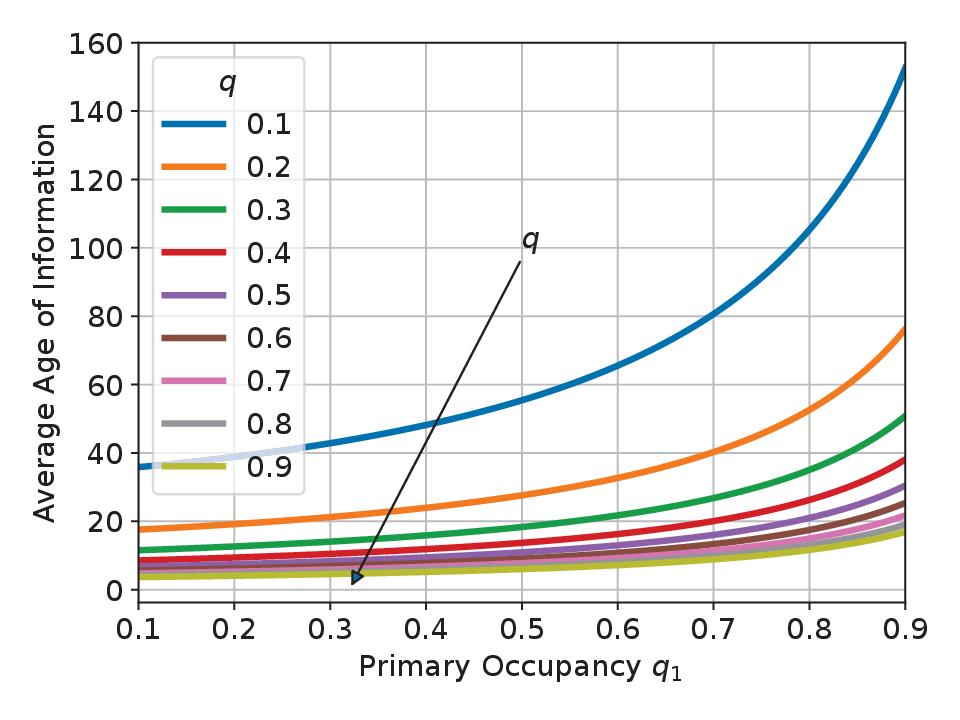}
     \setlength{\belowcaptionskip}{-12pt}
    \caption{Average AoI versus incumbent channel occupancy varying the probability that secondary has packets, $q$.}
    \label{fig:AAoI_q1}
\end{figure}

We consider $h(i,k)=1$ and $g(i,k)=2^{-4}$ to represent an attenuation coefficient over a distance. The SNR is assumed to be $\gamma = 0$ dB for the purposes of signal detection. We assume a CNN classifier is used for spectrum sensing by the transmitter, while the adversary uses a FNN classifier. Unless otherwise stated, we use packets of $64$ I/Q samples. When both the incumbent and the secondary user are transmitting, the jammer observes a $3$ dB increase in SNR and the detection accuracy is increased accordingly. 
Fig.~\ref{fig:AAoI_q1} shows the average AoI at the secondary receiver versus the probability of an incumbent transmission $q_1$. The presence of the incumbent should prevent the secondary user from transmitting, except for the misdetection events. As a result, the AoI increases with $q_1$. When incumbent occupancy is less than $50\%$, its effect on AoI is reduced. Also, the effect of a large $q_1$ is more prominent when the secondary transmission attempts are sparse. 

\begin{figure}
    \centering
    \vspace{-8pt}
    \includegraphics[width=0.95\columnwidth]{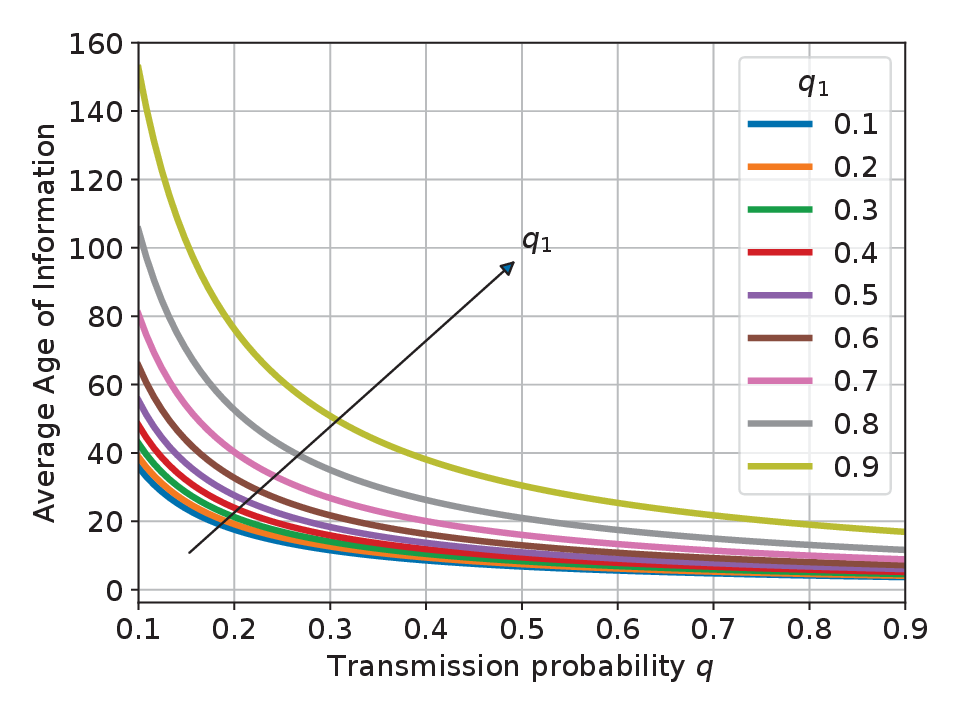}
     \setlength{\belowcaptionskip}{-18pt}
    \caption{Average AoI versus probability that secondary user intends to transmit (subject to spectrum sensing) varying $q_1$. }
    \label{fig:AAoI_q}
\end{figure}

Fig.~\ref{fig:AAoI_q} presents the average AoI versus the probability that the secondary node intends to transmit. Note that this is not the secondary transmit probability $q_2$, which also depends on $q_1$, but the independent variable $q$, which can be thought of as the rate that packets are generated at the transmitter. In the unconstrained case, more attempted transmissions translate into higher success probability and hence smaller AoI. If transmissions were constrained, by a power budget for example, these results indicate that limited gain is obtained when increasing the packet generation beyond $0.4$. Higher rates are particularly desirable when spectrum is scarce, meaning that the secondary user should take advantage of a transmission opportunity when one arrives. 

\begin{figure}
    \centering
    \includegraphics[width=0.95\columnwidth]{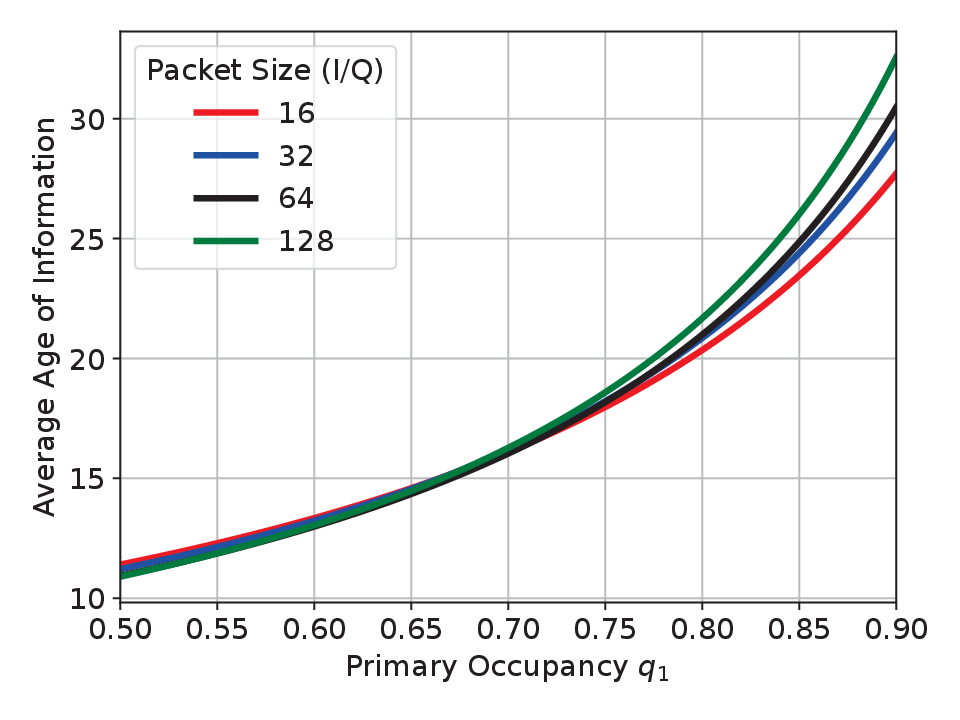}
    \setlength{\belowcaptionskip}{-18pt}
    \caption{Average AoI versus incumbent channel occupancy varying packet size (in I/Q samples) with $q=0.5$.}
    \label{fig:Pkt_q05}
\end{figure}

The effect of packet size is illustrated in Fig.~\ref{fig:Pkt_q05}, assuming the transmitter is activated with probability $q=0.5$, if the spectrum is classified to be available. Packet size is more relevant when spectrum resources are scarce, and the results suggest that small packets yield smaller AoI in that case, when $q_1$ is large. In this scenario, the small packets result in smaller detection probability by the jammer, hence reducing interference. On the other hand, when the probability of incumbent transmission is small, the behavior is reversed, and larger packet sizes result in smaller average AoI.

\begin{figure}
    \centering
    \vspace{-30pt}
    \includegraphics[width=\columnwidth]{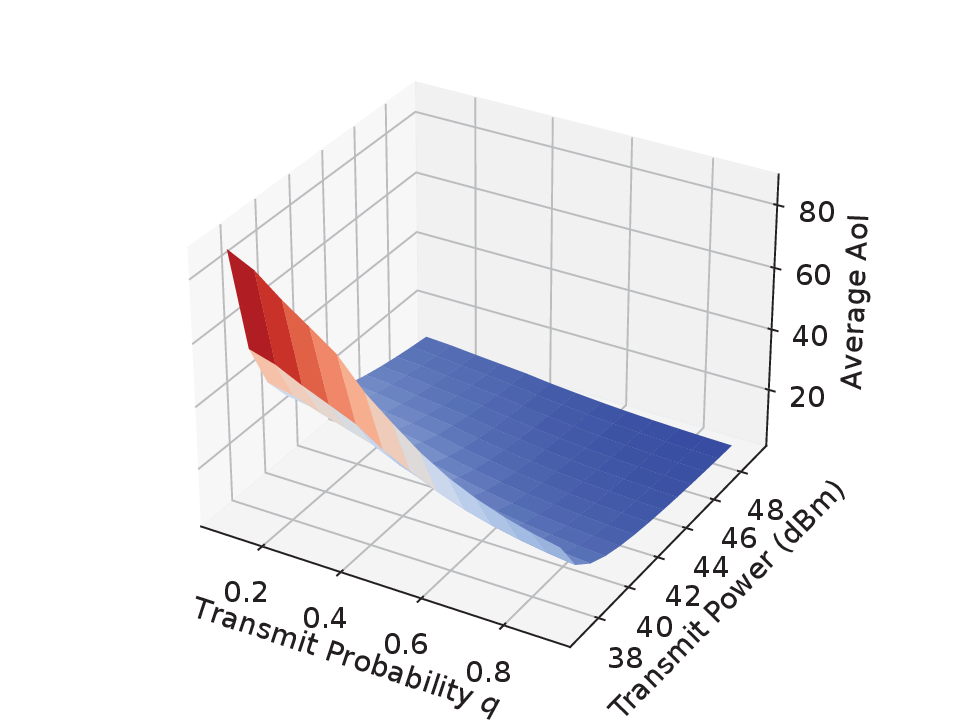}
    \setlength{\belowcaptionskip}{-18pt}
    \caption{Average AoI versus transmit power and transmit probability for secondary user.}
    \label{fig:AoI_PTX}
\end{figure}

The effect of power control is illustrated in Fig.~\ref{fig:AoI_PTX}, where we assume the incumbent occupies the channel with probability $q_1=0.5$ and show the average AoI with respect to the design parameters the secondary user can select. 

\begin{figure}
    \centering
    \vspace{-10pt}
    \includegraphics[width=\columnwidth]{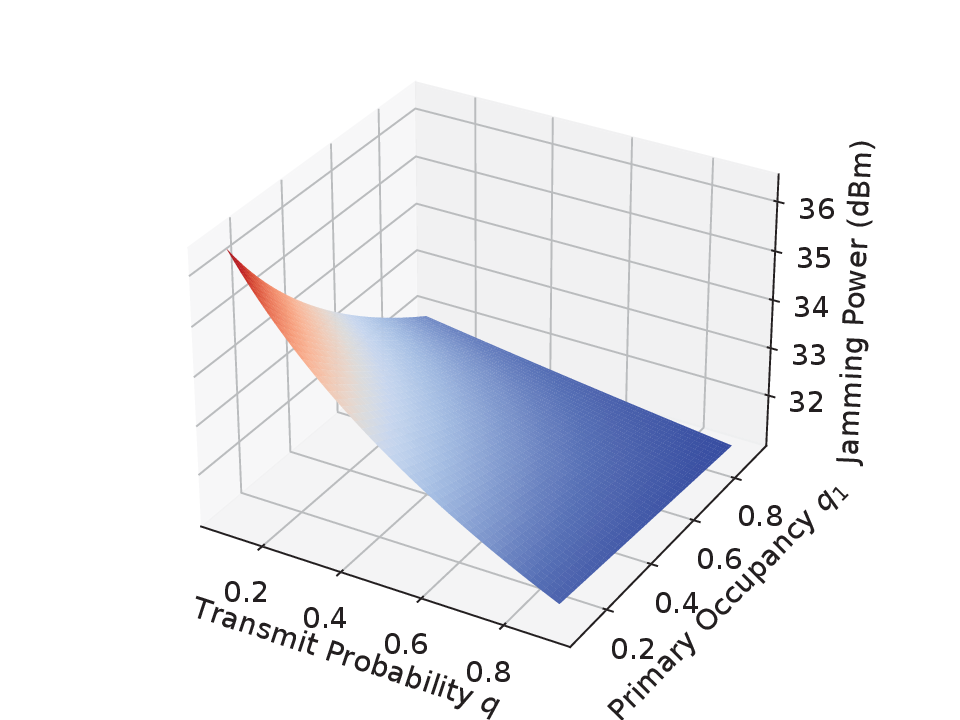}
    \setlength{\belowcaptionskip}{-18pt}
    \caption{Jamming power versus channel utilization.}
    \vspace{-10pt}
    \label{fig:PJ3D}
\end{figure}

Fig.~\ref{fig:PJ3D} shows the resulting jamming power as a function of channel occupancy by both users. Large transmission probabilities ($q$) result in more frequent jamming, which requires a reduction in jamming power used per time slot, given the power constraint at the adversary. To some extent, the spectrum sharing yields a reduced interference level from the perspective of each user, since the jammer will use some of its power to cause interference to the other user.

\section{Conclusion}\label{sec:conclusion}
We studied the intricate interplay between timeliness and security in NextG spectrum sharing in the presence of an adversary.
For this setting, we analyzed the effect of signal classification performance on the selection of jamming power for an adversary with limited resources and evaluated the average AoI. 
We showed the advantage of sharing the spectrum when the adversary has limited power budget and will be encouraged to reduce its jamming power with the increased occupancy of the channel. 
We also considered the effects of the wireless channels and design parameters as the packet size, and show that smaller packets can reduce the average AoI in a scenario of spectrum scarcity under the presence of an adversary. 
Our findings highlight pathways for developing resilient NextG communications protocols, emphasizing the connection between spectrum sharing, timeliness, and security.

\bibliographystyle{IEEEtran}
\bibliography{refs}

\end{document}